\documentclass[aps,prb,reprint,superscriptaddress,amsmath,amssymb]{revtex4-1}

\usepackage{graphicx}
\usepackage{epstopdf}
\usepackage{dcolumn}
\usepackage{bm}
\usepackage[version=3]{mhchem}
\usepackage{natbib,hyperref}
\hypersetup{colorlinks=true,linkcolor=blue,citecolor=blue,urlcolor=blue}
\usepackage{color}
\usepackage[normalem]{ulem}
\usepackage{float}
\usepackage{braket}
\usepackage{array}
\usepackage{bigstrut}
\usepackage{float}
\renewcommand{\emph}[1]{\textit{#1}}

\begin{document}

\title{Optical Stark metrology of CdSe quantum dots:\\
Reconciling the size-dependent oscillator strength with theory}

\author{Yanhao Tang}
\affiliation{Department of Physics and Astronomy, Michigan State University, East Lansing, MI 48824, USA}
\author{Mersedeh Saniepay}
\affiliation{Department of Chemistry, Michigan State University, East Lansing, MI 48824, USA}
\author{Chenjia Mi}
\affiliation{Department of Chemistry, Michigan State University, East Lansing, MI 48824, USA}
\author{R\'{e}mi Beaulac}
\affiliation{Department of Chemistry, Michigan State University, East Lansing, MI 48824, USA}
\author{John A. McGuire}
\email{mcguire@pa.msu.edu}
\affiliation{Department of Physics and Astronomy, Michigan State University, East Lansing, MI 48824, USA}

\date{\today}

\begin{abstract}
The $k\cdot p$ effective mass approximation (EMA) predicts large, nearly size-independent exciton oscillator strengths in quantum confined semiconductors. Yet, experimental reports have concluded that the total oscillator strength of the lowest-energy (1S$_{3/2}$1S$_{\text{e}}$) excitons in strongly confined CdSe NQDs is small and strongly size-dependent. Using the optical Stark effect, we show that the oscillator strength of the 1S$_{3/2}$1S$_{\text{e}}$ excitonic absorption peak in CdSe NQDs follows the predictions of the EMA. These oscillator strengths enable helicity-selective unsaturated Stark shifts corresponding to femtosecond pseudo-magnetic fields exceeding 100 T. 
\end{abstract}

\maketitle

Strong confinement in nanocrystal quantum dots (NQDs) has dramatic implications for fundamental physical processes, e.g., spin-carrier interactions\cite{Beaulac2009}, and applications. The most invoked and widely analyzed consequence of confinement is the size-dependence of optical transition energies. Equally important is the size-dependent oscillator strength, i.e., the light-matter interaction, of the lowest-energy transitions. Large oscillator strengths imply strong absorption and emission and so determine the performance of NQD-based photovoltaics\cite{StolleHarveyKorgel2013}, light-emitting diodes\cite{ShirasakiBawendiBulovic2013}, and bio-labels\cite{BruchezAlivisatos1998}. Despite debates over the appropriateness of the $k\cdot p$ effective-mass approximation (EMA) for calculations of the electronic structure of small NQDs\cite{Fu1997,Efros1998a,Fu1998a,Fu1998}, optical transition energies have been well reproduced by the EMA\cite{Ekimov1993,NorrisBawendi1996,Efros1996,Leatherdale2002, Yu2003,Jasieniak2009,CapekHens2010}. Meanwhile, one of the landmark achievements of EMA-based calculations of NQD electronic structure was the calculation of the exciton fine structure of CdSe NQDs, the most widely studied NQD system, and identification of the lowest-energy dark states\cite{Nirmal1995,Efros1996}. The EMA also predicts large, nearly size-independent values of the integrated oscillator strength of the lowest-energy excitons\cite{Brus1984,Hanamura1988,Kayanuma1988,Efros1996}. 

\begin{figure}[b!]
\includegraphics[width=\linewidth]{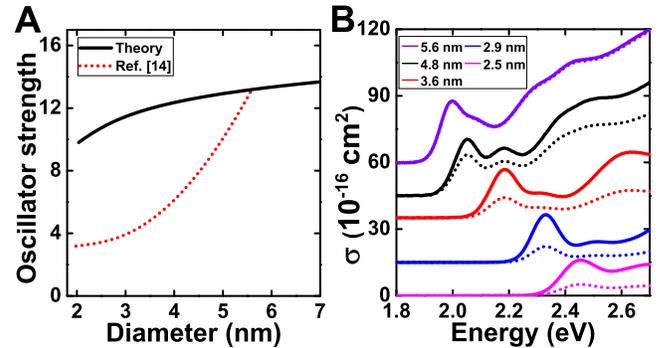}
\caption{\label{fig:abs_spec}\textbf{Comparisons between the effective-mass approximation and previously reported results.} (\textbf{A}) Total oscillator strength of the 1S$_{3/2}$1S$_{\text{e}}$ peak in CdSe NQDs based on the effective-mass approximation (EMA, black curve) and from a fit to experimental measurements in Ref.~\cite{Jasieniak2009} (dotted red curve). The EMA curve is based on a constant value of $(\hbar\omega f)_{1\text{S}_{3/2} 1\text{S}_\text{e}}$, which is taken equal to the sum of the orientationally averaged A and B excitons in the bulk ($f=14.3$, $\hbar\omega=1.84$ eV)~\cite{Voigt1979}. (\textbf{B}) Ground-state absorption cross section for NQDs of different sizes: Solid curves refer to the EMA, and dashed curves refer to the results of Ref.~\cite{Jasieniak2009}.}
\end{figure}

\begin{figure*}[t!]
	\includegraphics[width=\linewidth]{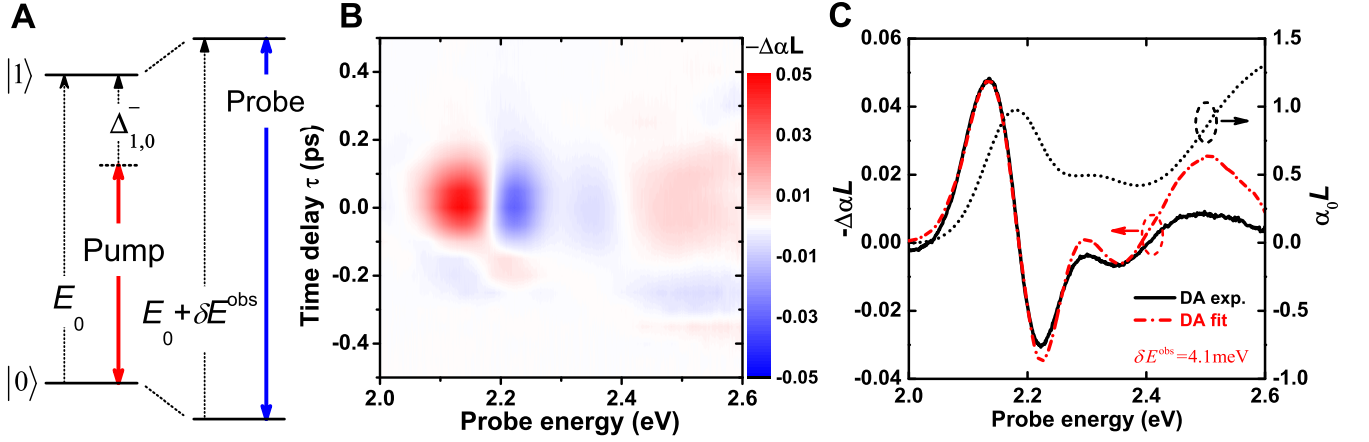}
	\caption{\label{fig:OSE_800}\textbf{Optical Stark effect for CdSe NQDs and co-linearly polarized Stark and probe fields.} (\textbf{A}) Illustration of the optical Stark effect showing the repulsion of two energy states $\ket{0}$ and $\ket{1}$ due to a red-detuned Stark pump. 
(\textbf{B}) Experimental differential absorption spectrum, $-\Delta\alpha\left(\hbar\omega,t\right) L=-\left[\alpha_{\text{pump}} \left(\hbar\omega,t\right)- \alpha_{0}\left(\hbar\omega\right)\right]L$, for 3.6~nm NQDs, where $\alpha_{\text{pump}} \left(\hbar\omega,t\right)$ and $\alpha_{0} \left(\hbar\omega\right)$ are the absorption coefficients of the NQD solution in the presence and absence, respectively, of the Stark pump and $L$ is the sample length. The Stark pump is at $E_p=1.55$~eV, corresponding to a -0.63~eV detuning from the 1S$_{3/2}$1S$_{\text{e}}$ absorption peak, with intensity $I_0=10.7\pm$1.1~GW~cm$^{-2}$. 
(\textbf{C}) $-\Delta\alpha(E,t=0) L$ (solid curve), $\alpha_0L$ (dotted curve), and $-\Delta\alpha_{\text{fit}}(E,0) L \equiv \left[\alpha_0\Big(E\Big) - \alpha_0\left(E+\delta E^{\text{obs}}_{1\text{S}_{3/2}1\text{S}_\text{e}}\right) \right]L$ (dashed-dotted curve) with $\delta E^{\text{obs}}_{1\text{S}_\text{e},1\text{S}_{3/2}}
= 4.1$~meV.}
\end{figure*}

As NQD radius decreases below the bulk exciton Bohr radius, $a_B^*$, the reduced number of unit cells comprising a NQD, and so contributing to the oscillator strength, is compensated by the increased volume of reciprocal-space contributing to the lowest-energy confined excitons (the 1S$_{3/2}$1S$_{\text{e}}$ excitons in CdSe\cite{Ekimov1993}). As a result, and as shown below, the EMA predicts the product of the energy ($\hbar\omega$) and the oscillator strength ($f$) to be size-independent for the 1S$_{3/2}$1S$_{\text{e}}$ exciton manifold\cite{Brus1984,Hanamura1988,Kayanuma1988,Efros1996}. Despite the aforementioned successes of the EMA, numerous measurements of CdSe NQDs based on challenging analytic estimates\cite{GongKelley2013} of NQD concentrations in solution suggest a strongly size-dependent value of $f_{\text{1S}_{3/2}\text{1S}_{\text{e}}}$ falling in small NQDs to $\sim$1/3 the bulk exciton value\cite{Klimov2000, Schmelz2001, Leatherdale2002, Yu2003, MelloDonegaKoole2009, Jasieniak2009, CapekHens2010}. The divergence between values of $f_{\text{1S}_{3/2}\text{1S}_{\text{e}}}$ determined by experiment and EMA calculations and the corresponding implications for the spectra of the ground-state absorption cross section are highlighted in Fig.~\ref{fig:abs_spec}. Notably, the experimental results have gone unexplained. If valid, these results imply a basic misunderstanding of the electronic structure of strongly confined NQDs and a failure of the EMA, while suggesting that single excitons in small NQDs are much less easily optically generated and manipulated than in larger NQDs or bulk\cite{Zhang2010}. 

Here, we use optical Stark metrology to obtain a measure of $f_{\text{1S}_{3/2}\text{1S}_{\text{e}}}$ that is free of estimates of NQD concentration and only weakly sensitive to the accuracy with which NQD size is known. We show the total oscillator strength of the 1S$_{3/2}\rightarrow$1S$_{\text{e}}$ transition in CdSe NQDs to be consistent with predictions of the EMA. (In the absence of explicit reference to the fine structure states, we refer to the manifold of 1S$_{3/2}\rightarrow$1S$_{\text{e}}$ fine structure transitions collectively as as the 1S$_{3/2}\rightarrow$1S$_{\text{e}}$ transition.) These large oscillator strengths enable helicity-selective, unsaturated Stark shifts of 17 meV corresponding to pseudo-magnetic fields exceeding 100 T and suggesting new possibilities for coherent optical spin manipulation in NQDs.

The OSE, illustrated in Fig.~\ref{fig:OSE_800}A, is a shift of an optical transition due to interaction with an optical field that transiently mixes the two states of the transition. The optical Stark shift (OSS) of states $i$ and $j$, of energies $E_i$ and $E_j>E_i$, connected by a dipole-allowed transition is given by second-order perturbation theory \cite{BruevichKhodovoi1968} as
\begin{equation}
\label{eq:OSS}
\delta E_j=-\delta E_i=\frac{1}{4} \left|\mathcal{E}_{\text{in}}\right|^2 \left|\mathbf{e}\cdot\vec{\mu}_{ji}\right|^2 \left(\frac{1}{\Delta_{ji}^-}+\frac{1}{\Delta_{ji}^+}\right),
\end{equation}
where $\mathbf{e}$ is the unit polarization vector of the electric field, $\vec{\mu}_{ji}\equiv -e\vec{r}_{ji}$ is the electric dipole transition matrix element, $e$ is the magnitude of the electron charge, $\Delta_{ji}^{\pm}=E_j-E_i\pm E_{p}$, and $E_{p}=\hbar\omega_p$ is the energy of the Stark pump. $\mathcal{E}_{\text{in}}$ is related to the pump intensity, $I_0$, by $|\mathcal{E}_{\text{in}}|^2 =2|F|^2I_0/(\epsilon_0n_{\text{s}} c)$. The OSS is an intrinsically single-exciton process, so the NQD concentration does not appear in Eq.~\ref{eq:OSS}. Since $I_0$ and $\Delta^{\pm}_{ji}$ are easily measured, the OSS reports directly on $|\mathbf{e}\cdot\vec{\mu}_{ji}|^2$, and consequently on the oscillator strength of the $i\rightarrow j$ transition:
\begin{equation}
\label{eq:f}
f_{ji}=\frac{2m_0\omega_{ji}}{\hbar e^2}\left< \left|\mathbf{e}\cdot\vec{\mu}_{ji} \right|^2\right>_{\Omega}
=\frac{2}{m_0\hbar\omega_{ji}}\left< \left|\mathbf{e}\cdot\vec{p}_{ji}\right|^2\right>_{\Omega},
\end{equation}
where $\vec{p}$ is the momentum operator and the angled brackets indicate an average over all orientations of the system (hereafter we drop the subscript $\Omega$). Since in the EMA $\vec{p}_{ji}$ is energy-independent for interband transitions, $\left(\hbar\omega f\right)_{\text{1S}_{3/2} \text{1S}_{\text{e}}}$ is predicted to be constant\cite{Brus1984,Efros1996}. 

In practice, the light-matter interaction in NQDs is often described in terms of the  absorption cross section. $f_{1\text{S}_{3/2}1\text{S}_{\text{e}}}$ is directly related to the energy-integrated absorption cross section of the 1S$_{3/2}$1S$_{\text{e}}$ peak\cite{DelerueLannoo2004}:
\begin{equation}
\label{eq:sigma1S}\bar{\sigma}_{1\text{S}_{3/2}1\text{S}_{\text{e}}}=\frac{\pi e^2 \hbar}{2\epsilon_0 n_{\text{s}} m_0 c} \left|F\left(\hbar \omega_{1\text{S}_{\text{e}},1\text{S}_{3/2}}\right)\right|^2 f_{1\text{S}_{3/2}1\text{S}_{\text{e}}},
\end{equation}
where $n_{\text{s}}$ is the solvent refractive index, $m_0$ is the electron mass in vacuum, $\epsilon_0$ is the permittivity of free space, $c$ is the vacuum speed of light, and $F=\mathcal{E}_{\text{in}}/ \mathcal{E}_{\text{out}}$ is the local field correction factor relating the electric field inside ($\mathcal{E}_{\text{in}}$) and outside ($\mathcal{E}_{\text{out}}$) the NQD.

Optical experiments were performed on CdSe NQDs with the wurtzite crystal structure and diameters $d=2.5$--6.7 nm (cf. $2a^*_{\text{B}}=11.2$~nm), which were synthesized by hot injection \cite{Yu2003,DuttaBeaulac2016a} or purchased from NN Labs. NQD diameters were determined from the energy of the 1S$_{3/2}$1S$_\text{e}$ absorption peak using the empirical sizing curve of Ref.~\onlinecite{Jasieniak2009}. NQDs were dissolved in toluene or, in the case of 6.7~nm dots, in CCl$_4$ and loaded into fused silica cuvettes with 1-mm solution path length. 

\begin{figure*}[t!]
	\includegraphics[width=\linewidth]
{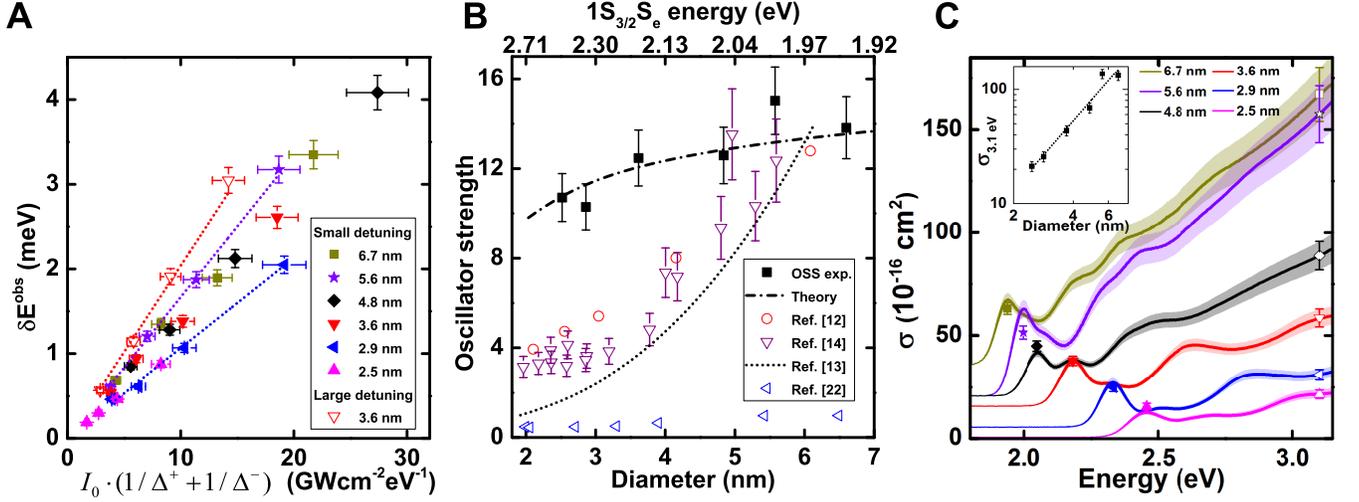}
	\caption{\label{fig:size_exc_dep} \textbf{Size-dependent optical Stark shift and oscillator strengths} (\textbf{A}) OSS for NQDs excited at 0.20--0.29~eV (filled symbols, see Table~\ref{tab:OSS} in Materials and Methods for detunings) or 0.63~eV (open triangles) below the 1S$_{3/2}$1S$_{\text{e}}$ peak. Dashed lines are linear fits to the data.  
(\textbf{B}) Energy-integrated oscillator strength of the 1S$_{3/2}$1S$_{\text{e}}$ peak as determined here (filled squares) and by previous approaches (open symbols and dotted line)\cite{Leatherdale2002,Yu2003,Jasieniak2009,Klimov2000}. The dash-dotted curve is the prediction of the effective-mass approximation for $\left(\hbar\omega f\right)_{\text{bulk}}=26.3$~eV, as shown in Fig. \ref{fig:abs_spec}A. 
(\textbf{C}) NQD ground-state absorption cross sections (solid curves). Absorbance spectra are measured by UV/vis absorption and scaled to the absorption cross section at 3.1~eV as determined by DA saturation measurements (open symbols). The spectra are offset by 5, 15, 20, 20, and $35\times10^{-16}$~cm$^2$ for NQDs of diameter 2.9, 3.6, 4.8, 5.6, and 6.7~nm, respectively. The broad bands represent the uncertainty from the DA saturation measurement. Solid points are the expected magnitude of the 1S$_{3/2}$1S$_{\text{e}}$ peak based on the oscillator strengths measured by the OSE. Inset shows the size-dependence of the absorption cross section at 3.1 eV (with the same units as in the main panel) and a power-law fit ($\sigma\left(3.1\,\text{eV}\right)\propto d^n$) with $n=2.0\pm0.2$.}
\end{figure*}

The OSE and carrier dynamics were measured by differential absorption (DA). For the OSE, we pumped samples with the 100-fs, 1.55-eV output of a 1-kHz Ti:sapphire laser (SpectraPhysics Spitfire PRO-XP) or the doubled output of a home-made optical parametric amplifier pumped by the laser. We generated real excited populations with the second harmonic of the laser fundamental. The probe was a supercontinuum produced by focusing $\sim$1~$\mu$J of the laser output onto a $c$-cut sapphire crystal and compressed for minimum dispersion around 2.2 eV with a pair of fused-silica prisms. Pump and probe beam diameters at the sample were respectively $\sim$1 mm and $\sim$0.1 mm. The angle between the pump and probe was 7.5$^{\circ}$. We measured the transmitted probe with 5 meV resolution using a CCD spectrometer synchronized to a mechanical chopper in the pump path. The peak pump fluence was determined by measurement of the pump power transmitted by a pinhole assuming a spatially uniform intensity over the pinhole. The peak intensity ($I_0$) was determined by a temporal Gaussian fit of $-\Delta\alpha(t)L$ at the lowest-energy DA peak. The linear (circular) polarization of pump and probe were controlled by half-wave (quarter-wave) plates. The pump power was adjusted by a set of neutral density filters. The solvent response was accounted for by subtracting the DA signal from the neat solvent under the same conditions as the NQDs. 

Fig.~\ref{fig:OSE_800}B shows a typical CdSe-NQD DA spectrum, -$\Delta\alpha\left(\hbar\omega,t\right) L$, where $\alpha$ and $L$ are respectively the absorption coefficient and length of the NQD solution. In Fig.~\ref{fig:OSE_800}C, we compare the DA spectrum at delay $t=0$ to the difference $-\Delta\alpha_{\text{fit}}L\equiv \left[\alpha_0(E) - \alpha_0(E+\delta E^{\text{obs}}_{1\text{S}_{\text{e}},1\text{S}_{3/2}})\right]L$, where the observed OSS, $\delta E^{\text{obs}}_{1\text{S}_{\text{e}},1\text{S}_{3/2}}$, is a fitting parameter used to match $\Delta\alpha_{\text{fit}}L$ to the amplitude of the measured $\Delta\alpha\left(E, t=0\right) L$ at the lowest-energy DA peak (see Appendix~\ref{app:obsOSS} for discussion of the relationship between the observed OSS of the 1S$_{3/2}$1S$_{\text{e}}$ peak and the actual OSS of the individual transitions within the spectral peak). 

The excitation- and size-dependence of the OSS for CdSe NQDs are shown in Fig.~\ref{fig:size_exc_dep}A for detuning $-\Delta^- \equiv -\Delta_{1\text{S}_{3/2}1\text{S}_{\text{e}}}^-=-0.20$ to $-0.29$~eV. The observed linear dependence of $\delta E^{\text{obs}}_{1\text{S}_{3/2} 1\text{S}_{\text{e}}}$ on $I_0(1/\Delta^- +1/\Delta^+)$ (the slope of the OSS data) varies with size %
only by about $\pm$20\%, immediately suggesting a similarly limited size-dependence of $f_{1\text{S}_{3/2}1\text{S}_\text{e}}$ in contrast to earlier results shown in Fig. \ref{fig:abs_spec}. Importantly, at equal values of $I_0\left(1/\Delta^+ + 1/\Delta^-\right)$ the OSS shown for 3.6~nm CdSe NQDs with a $-0.63$~eV (large) detuning yields a significantly different slope than at $-0.28$~eV (small) detuning, which is at odds with Eq.~\ref{eq:OSS}. This is evidence that the OSS of the 1S$_{3/2}$1S$_\text{e}$ peak is not determined solely by the interaction of light with the 1S$_{3/2}\rightarrow$1S$_\text{e}$ transition. To correctly determine $f_{1\text{S}_{3/2}1\text{S}_\text{e}}$, we must account for the contributions of other transitions to the OSS of the 1S$_{3/2}$1S$_\text{e}$ peak.

While an exact accounting of the OSS must address the excitonic (and biexcitonic) origins of the OSS\cite{Combescot1989}, for Stark-pump detunings large compared to the fine structure splittings and biexciton binding, the OSS is accurately calculated in a single-particle picture\cite{Combescot1990}. For the 1S$_{3/2}$1S$_{\text{e}}$ peak, this can be shown explicitly using detailed theories of single- and biexciton fine structure states \cite{Efros1996,Rodina2010} (see Appendix \ref{app:excitonOSS}). Nonetheless, we must still account for the OSS associated with all transitions involving the 1S$_{3/2}$ \emph{or} 1S$_\text{e}$ states and for the orientational distribution of NQDs (see Appendix~\ref{app:obsOSS}). For example, the oscillator strength of the 1S$_{\text{e}}\rightarrow$1P$_{\text{e}}$ transition is expected to be of the same order of magnitude as the 1S$_{3/2}\rightarrow$1S$_{\text{e}}$ transition\cite{GuyotSionnest1998} and so will contribute to the shift of the 1S$_{\text{e}}$ state and, hence, to the observed OSS of the 1S$_{3/2}$1S$_{\text{e}}$ peak to the extent that the detuning from the 1S$_{\text{e}}\rightarrow$1P$_{\text{e}}$ transition is not too large. The observed OSS is then determined by an interaction- and orientation-weighted average of the shift of each of the transitions constituting the 1S$_{3/2}$1S$_{\text{e}}$ peak: 
\begin{widetext}
\begin{equation}\label{eq:fullOSS}
\delta E^{\text{obs}}_{1\text{S}_{\text{e}}, 1\text{S}_{3/2}}
=\left<\displaystyle\sum_{\beta, M}\left|\bra{1\text{S}_{\text{e}}\beta}\mathbf{e}\cdot \hat{\bold{r}}\ket{1\text{S}_{\frac{3}{2}}M}_v \right|^2\right>^{-1}\left< \displaystyle\sum_{i,\beta, M} 
\left(\delta E^i_{1\text{S}_{\text{e}}\beta}-\delta E^i_{1\text{S}_{\frac{3}{2}}M}\right)
\left|\bra{1\text{S}_{\text{e}}\beta}\mathbf{e}\cdot \hat{\bold{r}}\ket{1\text{S}_{\frac{3}{2}}M}_v\right|^2
\right>
,
\end{equation}
\end{widetext}
where $\delta E^i_j$ indicates the OSS of level $j$ due to the $i\rightarrow j$ transition. To highlight the degree to which the OSS of the 1S$_{3/2}$1S$_{\text{e}}$ peak is due to the 1S$_{3/2}\rightarrow$1S$_{\text{e}}$ transitions or other transitions involving the 1S$_{3/2}$ or 1S$_{\text{e}}$ states, we can formally write Eq.~\ref{eq:fullOSS} as in Eq.~\ref{eq:OSS} via a NQD-diameter- and pump-energy-dependent factor $\gamma=\gamma(d,E_p)$: 
\begin{align}
\label{eq:OSS1S}
\delta E^{\text{obs}}_{1\text{S}_{\text{e}},1\text{S}_{\frac{3}{2}}}=&\gamma(d,E_p) \frac{\left|F\right|^2 }{\epsilon_0 n_{\text{s}} c}I_0\left(\frac{1}{\Delta^-}+\frac{1}{\Delta^+}\right)\nonumber\\
&\times\sum_{\beta, M}\left< \left|\mathbf{e}\cdot
\vec{\mu}_{1\text{S}_{\text{e}}\beta,1\text{S}_{\frac{3}{2}}M} \right|^2\right>,
\end{align}
where $\beta$ ($M$) is the projection of the electron (hole) angular momentum along the NQD $c$ axis.  $f_{1\text{S}_{3/2}1\text{S}_\text{e}}$ can then be related to the observed OSS by 
\begin{align}
\label{eq:f1S}
f_{1\text{S}_{3/2}1\text{S}_\text{e}}&= \frac{2m_0\omega_{1\text{S}_\text{e},1\text{S}_{3/2}}}{\hbar e^2}\sum_{\beta, M}\left< \left|\mathbf{e}\cdot
\vec{\mu}_{1\text{S}_{\text{e}}\beta,1\text{S}_{3/2}M}\right|^2 \right>
\nonumber\\
&=\frac{2 \epsilon_0 n_{\text{s}} c m_0 \omega_{1\text{S}_\text{e},1\text{S}_{3/2}}}{\gamma(d,E_p)\hbar e^2 \left| F \right|^2} \frac{\delta E^{\text{obs}}_{1\text{S}_{\text{e}},1\text{S}_{3/2}}}
{I_0\left(\frac{1}{\Delta^-}+\frac{1}{\Delta^+}\right)}.
\end{align}

The key computational parameter in Eqs.~\ref{eq:OSS1S} and \ref{eq:f1S} is $\gamma(d,E_p)$. As shown in Appendices~\ref{app:obsOSS} and \ref{app:excitonOSS}, accounting solely for interactions between the pump and the 1S$_{\text{e}}\rightarrow$1S$_{3/2}$ transitions, $\gamma(d,E_p)$ is exactly 2/5 in a single-particle picture, while variations due to excitonic effects are $<4$\% when pump detunings are large compared to the exciton fine structure splittings and biexciton binding. When including other transitions to or from the 1S$_{\text{e}}$ or 1S$_{3/2}$ states, for detunings of $-0.20$ to $-0.29$~eV and $d=2.5$ to 6.7~nm, the EMA yields $\gamma(d,E_p)=0.60$--0.66 (details in Table~\ref{tab:OSS} of Appendix~\ref{app:obsOSS}). 

The calculated $\gamma(d,E_p)$ and measured $\delta E^{\text{obs}}_{1\text{S}_{\text{e}},1\text{S}_{3/2}}$ yield the 1S$_{3/2}$1S$_\text{e}$ oscillator strengths shown in Fig.~\ref{fig:size_exc_dep}B, where we also show previous estimates of $f_{1\text{S}_{3/2}1\text{S}_{\text{e}}}$.  Most importantly, the values of
$f_{1\text{S}_{3/2}1\text{S}_{\text{e}}}$ measured here closely match theory: the energy-integrated oscillator strength of the 1S$_{3/2}$1S$_{\text{e}}$ peak in CdSe depends only weakly on size. Although $\gamma(d,E_p)$ is markedly different at $E_{p}=1.55$~eV (detunings of $-0.4$ to $-0.9$~eV) than at smaller detunings, the resulting $f_{1\text{S}_{3/2}1\text{S}_{\text{e}}}$ are the same as in Fig.~\ref{fig:size_exc_dep}B (see Fig.~\ref{fig:slopes} in Appendix~\ref{app:sat}); this consistency confirms the validity of the approach.  $f_{1\text{S}_{3/2}1\text{S}_{\text{e}}}$ drops from $\sim$14 in the largest dots to $\sim$10 in the smallest, while $\left(\hbar \omega f\right)_{1\text{S}_{3/2}1\text{S}_{\text{e}}}= 27\pm2$~eV, at least three times larger than previous estimates for the smallest NQDs\cite{Klimov2000, Schmelz2001, Leatherdale2002, Yu2003, MelloDonegaKoole2009, Jasieniak2009, CapekHens2010}. For comparison, the orientationally averaged sum of A and B exciton oscillator strengths per CdSe unit cell in the bulk is $f_{\text{unit}}\sim2.2\times10^{-3}$ \,\cite{Voigt1979},  which yields a combined oscillator strength of $f_{\text{X}}=f_{\text{unit}} V_{\text{X}}/V_{\text{unit}}\sim14$.  The measured OSS for CdTe NQDs (see Fig.~\ref{fig:slopes}A in Appendix~\ref{app:sat}) is the same as for CdSe NQDs of similar size and detuning, as expected given the similar electronic structure of both systems\cite{Efros1998}. Notably, the values of $f_{1\text{S}_{3/2}1\text{S}_{\text{e}}}$ found here for CdSe NQDs are also similar to those reported for CdTe NQDs\cite{KamalHens2012} as expected given the similar electronic parameters (gap and effective masses) of bulk CdSe and CdTe. 

As a further consistency check on $f_{1\text{S}_{3/2}1\text{S}_{\text{e}}}$, we measure $\sigma\left(3.1\,\text{eV}\right)$, the absorption cross section per dot at 3.1~eV, by DA saturation of the 1S$_{3/2}$1S$_{\text{e}}$ transition under 3.1~eV excitation (see Figs.~\ref{fig:TA_dynamics} and \ref{fig:sat_abs} in Appendix~\ref{app:sat})\cite{Klimov2000}. As shown in Fig.~\ref{fig:size_exc_dep}C, the absorption cross section at the peak of the 1S$_{3/2}$1S$_{\text{e}}$ absorption feature determined from the OSS is in close agreement with the absorption cross section determined by DA saturation, again supporting the accuracy of our approach. A power-law fit of the diameter-dependence of $\sigma\left(3.1\,\text{eV}\right)$ in the inset of Fig.~\ref{fig:size_exc_dep}C reveals a $d^{2.0\pm0.2}$ dependence. By comparing UV/vis spectra at 3.1 and 3.5 eV, we find that the same quadratic dependence holds at 3.5 eV. This observation is in contrast to earlier assumptions\cite{Klimov2000,MelloDonegaKoole2009} and reports\cite{Leatherdale2002,Jasieniak2009} of a $d^3$ size-dependence, as would be expected when confinement is irrelevant. However, Hens and collaborators have shown that for CdSe and CdTe the absorption spectra are influenced by confinement even at 3.5~eV\cite{CapekHens2010,KamalHens2012}, which makes assumptions of a $d^3$ dependence of $\sigma\left(3.5\,\text{eV}\right)$ questionable. Our observation can be qualitatively understood as a result of quantum confinement: as the NQD diameter increases, the energy spacing between different transitions increases, leading to a $d^2$-dependent density of transitions in the high-energy regime. Notably, the ratio of $\sigma\left(3.1\,\text{eV}\right)$ to $f_{1\text{S}_{3/2}1\text{S}_{\text{e}}}$ measured here shows a quadratic size-dependence, consistent with previous studies showing a $d^2$ dependence of the ratio of the high-energy absorption cross section to $f_{1\text{S}_{3/2} 1\text{S}_{\text{e}}}$\cite{Jasieniak2009, Klimov2000,MelloDonegaKoole2009,Leatherdale2002}. 

The discrepancies, reflected in Fig.~\ref{fig:size_exc_dep}B, with earlier experimental reports of $\sigma_{1\text{S}_{3/2}1\text{S}_\text{e}}$ and $f_{1\text{S}_{3/2}1\text{S}_\text{e}}$ for CdSe NQDs may be partly explained by the sensitive dependence of prior analytic approaches on accurate determination of NQD concentrations, which typically rely on assumptions about, e.g., shape, stoichiometry, distribution of stoichiometric excess, and reaction yield and are extremely sensitive to the accuracy of measurements of NQD diameter \cite{GongKelley2013}. For example, if the radii of small NQDs were underestimated by one unit cell, correction would shift the results of Refs.~\onlinecite{Leatherdale2002} and \onlinecite{Jasieniak2009} (shown in Fig.~\ref{fig:size_exc_dep}B) into agreement with the present results. Although such a large measurement error seems unlikely, this example highlights how sensitive the analytic approach is to the underlying measurements and assumptions. The light-matter interaction has also been addressed by PL lifetime measurements in CdTe and CdSe \cite{vanDriel2005, MelloDonegaKoole2009,GongKelley2013}, but nonradiative decay processes and size-dependent fine structure make it difficult to extract the intrinsic $f_{1\text{S}_{3/2}1\text{S}_\text{e}}$ from CdSe by PL lifetime measurements. In contrast to traditional analytic approaches, the oscillator strength determined by the OSS at small detunings does not require knowledge of the NQD concentration and, according to the EMA, is relatively insensitive to experimental estimates of NQD size: the OSS of the 1S$_{3/2}$1S$_{\text{e}}$ peak is dominated by the 1S$_{3/2}\rightarrow$1S$_{\text{e}}$ transition, so that $\gamma(d,E_p)$ in Eq.~\ref{eq:OSS1S} is calculated to vary across the entire size range studied here by only about 10\% for the small-detuning data of Fig.~\ref{fig:size_exc_dep}. Likewise, 
for Stark pump detunings that are large compared to the unresolved features of the exciton fine structure, the observed OSS is only weakly sensitive to the size-dependent exciton fine structure.

\begin{figure}[t!]
\centering
	\includegraphics[width=\linewidth]{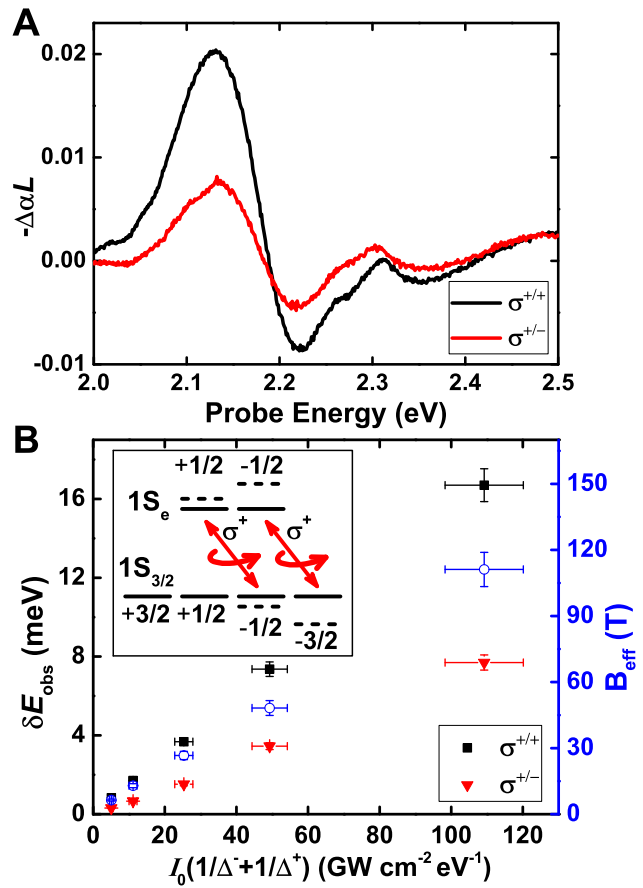}
	\caption{\label{fig:cir_exc}\textbf{The OSE for circularly polarized excitation.} (\textbf{A}) $-\Delta\alpha L$ spectra for 3.6~nm CdSe NQDs at $\tau=0$ ps for co- ($\sigma^{+/+}$) and counter-circularly ($\sigma^{+/-}$) polarized Stark field and probe with $E_p=1.904$~eV and $I_0=2.8\pm$0.3~GW/cm$^2$. (\textbf{B}) The OSS at $\tau=0$ ps for co- (black squares) and counter-circularly (red triangles) polarized pump and probe are shown as a function of $I_0(1/\Delta^-+1/\Delta^+)$. The corresponding pseudo-magnetic field, $B_{\text{eff}}$ is shown by empty blue circles. The inset is a schematic diagram of the OSS of the 1S$_{3/2}$ and 1S$_{\text{e}}$ states for right circularly polarized pump.}
\end{figure}

The oscillator strength determines the ease of coherent optical manipulation of carriers and spins. Hence, large oscillator strengths underpin proposals for quantum information processing in self-assembled quantum dots\cite{Imamoglu1999, Chen2004, PryorFlatte2006}. Our measured oscillator strengths suggest the potential for a large helicity-selective OSS. Fig.~\ref{fig:cir_exc}A shows the OSS of 3.6~nm NQDs for a 1.904~eV Stark field (-0.28~eV detuning) and co- and counter-circular polarization. As shown in Fig.~\ref{fig:cir_exc}B, the difference in the OSS for opposite helicities increases linearly with pump fluence and reaches 9~meV. This corresponds to a pseudo-magnetic field $B_{\text{eff}}=\left(\delta E^{+}-\delta E^{-}\right)/\left(\mu_B g_{\text{eff}}\right)=110$~T, where $\delta E^{+(-)}$ is the OSS under co(counter)-circularly polarized Stark and probe fields, $\mu_B$ is the Bohr magneton, and $g_{\text{eff}}=1.4$\cite{kuno1998}. For a pump pulse of length $\tau=100$~fs, this corresponds to a tipping angle $\theta\approx \left(\delta E^{+}-\delta E^{-}\right)\tau/\hbar=1.4$, or nearly $\pi/2$, similar to that observed in metal-semiconductor colloidal hetero-nanostructures\cite{Zhang2010}. Notably, at the highest fluences in Fig.~\ref{fig:cir_exc}B the OSS reaches 17~meV while still in the linear regime; no saturation is observed. Measurement of larger shifts was limited by contributions to the DA signal from carriers generated by two-photon absorption. Previous measurements of the OSE in NQDs grown in glass matrices\cite{Tsuda1996,GuptaPRB2001} or as hybrid metal-semiconductor heterostructures\cite{Zhang2010} saw saturation of the OSS at shifts of 1 -- 15~meV. The larger unsaturated OSS observed here may be a consequence of larger detunings and correspondingly reduced generation of real populations of screening carriers than in earlier studies or of reduced two-photon absorption in the substantially smaller NQDs studied here than in Refs.~\onlinecite{Tsuda1996} and \onlinecite{Zhang2010}. Larger tipping angles may be attainable by tuning the pump to minimize the ratio of two-photon absorption to the OSS and allow greater Stark pump intensities.

The long-standing puzzle over the size-dependence of the 1S$_{3/2}$1S$_{\text{e}}$ oscillator strength in CdSe NQDs highlights the challenges of determining fundamental electronic properties of even the nominally best understood NQD materials. The optical Stark effect offers a general approach for measuring the oscillator strengths in a wider variety of strongly confined systems, such as heterostructured and wide-band-gap NQDs, than is readily achieved by traditional analytical approaches. In CdSe NQDs, the optical Stark effect reveals that, despite long-standing experimental reports to the contrary, the EMA correctly accounts for the oscillator strength of the lowest-energy excitons. At the same time, the demonstrated generation of large optical Stark shifts in the absence of coupling to plasmonic resonances\cite{Zhang2010} allows for expanded possibilities for coherent manipulation of excitons in NQDs.

\begin{acknowledgments}
We thank Carlo Piermarocchi for helpful discussions.
\end{acknowledgments}

\appendix

\section*{Appendices}
We present details of calculations of the optical Stark shift (OSS) of the 1S$_{3/2}$1S$_{\textrm{e}}$ absorption peak for Stark pump detunings that are large compared to the splittings of the exciton fine structure and biexciton binding. We show that these calculations yield the same oscillator strengths for experimental OSS measurements performed both at small detunings, for which the 1S$_{3/2}\rightarrow$1S$_{\textrm{e}}$ transition is the dominant contribution to the OSS of the 1S$_{3/2}$1S$_{\textrm{e}}$ absorption peak, and at large detunings, for which the transitions other than the 1S$_{3/2}\rightarrow$1S$_{\textrm{e}}$ transition account for most of the OSS of the 1S$_{3/2}$1S$_{\textrm{e}}$ peak. Using an excitonic picture, we also show that, for detunings that are large compared to the splittings of the exciton fine structure, the size-dependence of the exciton fine structure has little impact on the OSS of the 1S$_{3/2}$1S$_{\textrm{e}}$ absorption peak: a calculation of the observed OSS based on a single-particle picture yields a result differing $<$4\% from a calculation based on an excitonic picture. Finally, we show an example of the differential absorption (DA) dynamics traces for the 1S$_{3/2}$1S$_{\textrm{e}}$ peak after 3.1~eV excitation and the DA saturation data for the nanocrystal quantum dots (NQDs) that we have studied. 

\section{\label{app:obsOSS}Calculation of the observed Stark shift} 
The OSS of the 1S$_{3/2}$1S$_{\text{e}}$ peak is due to the collective shift of individual fine structure transitions constituting the peak. To determine the oscillator strength, $f_{1\text{S}_{3/2}1\text{S}_{\text{e}}}$, from the observed OSS, we first note that, using single-particle notation, the absorption coefficient of the Stark shifted 1S$_{3/2}$1S$_{\text{e}}$ peak ($\alpha^{\prime}_{1\text{S}_{3/2}1\text{S}_{\text{e}}}$) is given by
\begin{align*}
\alpha^{\prime}_{1\text{S}_{3/2} 1\text{S}_{\text{e}}}\left(\hbar\omega\right) &\propto  
\left< \displaystyle\sum_{\substack{\beta=\uparrow,\downarrow\\ M=\pm 3/2,\pm1/2}} g\left(\hbar\omega-
\left[E_{\beta,M}-\delta E_{\beta,M}\right] \right)\right. \\&\quad\times\left. \vphantom{\displaystyle\sum_{}} 
\left|\bra{1\text{S}_{\text{e}}\beta}\mathbf{e}\cdot \hat{\bm{r}}\ket{1\text{S}_{3/2}M}_v\right|^2 \right>,\end{align*}
where $E_{\beta,M}\equiv E_{1\text{S}_{\text{e}}\beta}-E_{1\text{S}_{3/2}M}$,
\begin{equation*}
\delta E_{\beta,M}\equiv\displaystyle\sum_{i}\left(\delta E^i_{1\text{S}_{\text{e}}\beta}-\delta E^i_{1\text{S}_{3/2}M} \right),
\end{equation*} 
$\delta E^i_j$ is the orientation-dependent OSS of state $j$ due to the $i\rightarrow j$ transition, $g(\hbar\omega-E_{j,i})$ is the absorption line shape of the $i\rightarrow j$ transition, $\beta$ and $M$ are respectively the projections of the electron spin and hole angular momenta onto the NQD $c$ axis, and the outer, angled brackets indicate an angular average over the NQD orientational distribution. The subscript $v$ indicates the state of a valence-band electron ($M_v=-M_{\text{hole}}$).  (Note that besides the use just described in $\delta E^i_j,$ superscripts are used in two other ways in this manuscript. Superscripts $c$ and $v$ in $\delta E^{c(v)}_j$ indicate that state $j$ is a state of the conduction or valence band, respectively, while $\delta E^{\text{obs}}_{j,i}$ always refers to an observed change in the $i\rightarrow j$ transition.) Taking each transition in the 1S$_{3/2}$1S$_{\text{e}}$ manifold as characterized by the same line shape and the OSS for the transitions as small compared to the linewidth, Taylor expansion yields 
\begin{align*}
\Delta\alpha_{1\text{S}_{3/2} 1\text{S}_{\text{e}}}
\left(\hbar\omega\right) 
&
\equiv 
\left[\alpha^{\prime}_{1\text{S}_{3/2} 1\text{S}_{\text{e}}} \left(\hbar\omega\right) -
\alpha_{1\text{S}_{3/2} 1\text{S}_{\text{e}}} 
\left(\hbar\omega\right)\right] \\
& \propto \displaystyle\sum_{\beta, M} \left[\left.\frac{d g}{d\left(\delta E_{\beta,M}\right)}\right|_{\delta E_{\beta,M}=0}\right] \\
&\quad\times 
\left<\delta E_{\beta,M} \vphantom{\displaystyle\sum_{\alpha, M}} \left| \bra{1\text{S}_{\text{e}}\beta}\mathbf{e}\cdot \hat{\bm{r}}\ket{1S_{3/2}M}_v \right|^2 \right>.
\end{align*}
For linewidths much larger than the fine structure splittings, the quantity in square brackets can be taken as independent of the particular fine structure transition, and the differential absorption coefficient becomes
\begin{align*}
\Delta\alpha_{1\text{S}_{3/2} 1\text{S}_{\text{e}}}
\left(\hbar\omega\right) &\propto \displaystyle\sum_{\beta, M} \left<\delta E_{\beta,M} \vphantom{\displaystyle\sum_{\alpha, M}} \left| \bra{1\text{S}_{\text{e}}\beta}\mathbf{e}\cdot \hat{\bold{r}}\ket{1\text{S}_{3/2}M}_v \right|^2 \right>.\end{align*}
The \emph{observed} optical Stark shift ($\delta E^{\text{obs}}_{1\text{S}_{\text{e}}, 1\text{S}_{3/2}}$) is defined as the uniform shift of all transitions in the 1S$_{3/2}$1S$_{\text{e}}$ manifold that yields the same value of  $\Delta\alpha_{1\text{S}_{3/2} 1\text{S}_{\text{e}}} \left(\hbar\omega\right)$ as measured in DA, so that 
$\delta E^{\text{obs}}_{1\text{S}_{\text{e}}, 1\text{S}_{3/2}}$ satisfies
\begin{widetext}
\begin{align*}
\Delta\alpha_{1\text{S}_{3/2} 1\text{S}_{\text{e}}}
\left(\hbar\omega\right) \propto  &\,
\delta E^{\text{obs}}_{1\text{S}_{\text{e}}, 1\text{S}_{3/2}}
\displaystyle\sum_{\beta, M} \left< \vphantom{\displaystyle\sum_{\alpha, M}} \left| \bra{1\text{S}_{\text{e}}\beta}\mathbf{e}\cdot \hat{\bold{r}}\ket{1\text{S}_{3/2}M}_v \right|^2 \right>,\end{align*}
or equivalently
\begin{align*}
\delta E^{\text{obs}}_{1\text{S}_{\text{e}}, 1\text{S}_{3/2}}
&=\left<\displaystyle\sum_{\beta, M}\left|\bra{1\text{S}_{\text{e}}\beta}\mathbf{e}\cdot \hat{\bm{r}}\ket{1\text{S}_{\frac{3}{2}}M}_v \right|^2\right>^{-1}
\left< \displaystyle\sum_{i,\beta, M} 
\left(\delta E^i_{1\text{S}_{\text{e}}\beta}-\delta E^i_{1\text{S}_{\frac{3}{2}}M}\right)
\left|\bra{1\text{S}_{\text{e}}\beta}\mathbf{e}\cdot \hat{\bm{r}}\ket{1\text{S}_{3/2}M}_v\right|^2
\right>,
\end{align*}
\end{widetext}
which is Eq.~\ref{eq:fullOSS}. 

From the right side of Eq.~\ref{eq:fullOSS}, we pull out a factor of 
\begin{equation*}
\frac{|F|^2}{\epsilon_0 n_{\text{s}} c} I_0 
\tilde{\Delta}_{1\text{S}_{\text{e}}, 1\text{S}_{3/2}}^{-1}
\left<\sum_{\beta,M} \left|\bra{1\text{S}_{\text{e}}\beta}\mathbf{e}\cdot \hat{\boldsymbol{\mu}}\ket{1\text{S}_{\frac{3}{2}}M}_v \right|^2\right>,
\end{equation*}
where $\tilde{\Delta}^{-1}_{1\text{S}_{\text{e}}, 1\text{S}_{3/2}}\equiv 1/\Delta^-_{1\text{S}_{\text{e}}, 1\text{S}_{3/2}}+1/\Delta^+_{1\text{S}_{\text{e}}, 1\text{S}_{3/2}}$. We write what remains as a size- and Stark-pump-energy-dependent factor $\gamma(d,E_p)$:
\begin{align*}
\delta E^{\text{obs}}_{1\text{S}_{\text{e}},1\text{S}_{3/2}} =&
\gamma(d,E_p) \frac{|F|^2}{\epsilon_0 n_{\text{s}} c} I_0 
\tilde{\Delta}_{1\text{S}_{\text{e}}, 1\text{S}_{3/2}}^{-1}%
\nonumber\\
&\times
\left<\sum_{\beta,M} \left|\bra{1\text{S}_{\text{e}}\beta}\mathbf{e}\cdot \hat{\boldsymbol{\mu}}\ket{1\text{S}_{3/2}M}_v \right|^2\right>,
\end{align*}
which is Eq.~\ref{eq:OSS1S}. This is simply a convenient way of characterizing the relative contributions of different transitions to the OSS of the 1S$_{3/2}$1S$_{\text{e}}$ absorption peak.

We calculate $\delta E^{\text{obs}}_{1\text{S}_{\text{e}}, 1\text{S}_{3/2}}$ using the effective mass approximation for the single-particle states. For clarity in the following, we reproduce the essential elements of the treatment by Ekimov et al.\cite{Ekimov1993} For spherical NQDs, the electron states are labeled in terms of the orbital angular momentum, $l$; its projection along the $c$ axis, $m$; and the spin $\beta$:
\begin{equation}
\label{eq:electrons}
\Psi^c_{nlm\beta}(\vec{r})=A_{ln}Y_l^m(\theta,\phi)j_l(k_{ln}r)u^c_{\beta},
\end{equation}
where $A_{ln}$ is a normalization constant, $Y_l^m(\theta,\phi)$ are spherical harmonics, $j_l(k_{ln} r)$ is the spherical Bessel function, $k_{ln}$ is the $n^{th}$ solution of the eigenvalue problem for the states with angular momentum $l$, and $u^c_{\beta}$ is the conduction-band-edge Bloch function. For an infinite potential, $j_l(k_{ln}r)=0$ at $r=a$, where $a$ is the radius of the NQD. However, we follow Norris and Bawendi in employing a finite electron confinement potential for which the electron wave functions outside the NQD take a similar form as above but with a different wave vector, $k_m$, in the matrix surrounding the NQD\cite{NorrisBawendi1996}. The values of the eigenenergies and consequently $k_c$ and $k_m$ are determined by solution of the boundary condition expressed by Eq.~1 of Ref.~\onlinecite{NorrisBawendi1996}.

The hole states are more complicated than the electron states due to the multiple valence bands. The hole states are expressed in terms of the total angular momentum,  $N$, which is the sum of the angular momenta of the Bloch and envelope wave functions, and its projection along the $c$ axis, $M$:
\begin{widetext}
\begin{align}
\label{eq:holes}
\Psi_{N,M}^{v,\pm}&(\vec{r})=\sqrt{2N+1}\left[
\displaystyle\sum_{l^{\pm}=N\pm\frac{1}{2},N\mp\frac{3}{2}}(-1)^{l-\frac{3}{2}+M}R_l^{\pm}(r) 
\displaystyle\sum_{\mu=-3/2}^{3/2}\left(\begin{array}{c c c}
  l & \frac{3}{2} & N \\ 
  m & \mu & -M
 \end{array}\right)Y_l^m(\theta,\phi)u^v_{3/2,\mu} \right. \nonumber \\
&\qquad\left. +(-1)^{N\pm1/2-1/2+M}R_s^{\pm}(r)
\times\displaystyle\sum_{\mu=-1/2}^{1/2}\left(\begin{array}{c c c}
  N\pm\frac{1}{2} & \frac{1}{2} & N \\ 
  m & \mu & -M
 \end{array}\right)Y_{N\pm\frac{1}{2}}^m(\theta,\phi)u^v_{1/2,\mu}\right],
\end{align}
\end{widetext}
where $m+\mu=M$, the $R_{l(s)}^{\pm}(r)$ are radial envelopes for the $J=3/2$ ($J=1/2$) holes, the $2\times3$ arrays are the Wigner 3$j$ symbols, $u^v_{3/2,\mu}$ are the zone-center Bloch functions of the heavy and light hole bands, $u^v_{1/2,\mu}$ are the zone-center Bloch functions of the split-off band, and the superscript $\pm$ refers to the parity of the wave function. The conduction- and valence-electron Bloch functions are
\begin{widetext}
\begin{align*}
u^c_{\uparrow}=S\uparrow, & \quad u^c_{\downarrow}=S\downarrow,\\
u^v_{3/2,3/2}=\frac{1}{\sqrt{2}}(X+iY)\uparrow, &
\quad u^v_{3/2,-3/2}=\frac{i}{\sqrt{2}}(X-iY)\downarrow, \\
u^v_{3/2,1/2}=\frac{i}{\sqrt{6}}\left[(X+iY)\downarrow - 2Z\uparrow\right], & \quad  
u^v_{3/2,-1/2}=\frac{1}{\sqrt{6}}\left[(X-iY)\uparrow + 2Z\downarrow\right], \\
u^v_{1/2,1/2}=\frac{1}{\sqrt{3}}\left[(X+iY)\downarrow + Z\uparrow\right], & \quad 
u^v_{1/2,-1/2}=\frac{i}{\sqrt{3}}\left[-(X-iY)\uparrow + Z\downarrow\right]. \\
\end{align*}
\end{widetext}

We first calculate $\delta E^{\text{obs}}_{1\text{S}_{\text{e}}, 1\text{S}_{3/2}}$ in the case that we neglect all transitions except the 1S$_{3/2}\rightarrow$1S$_{\text{e}}$ transition. The OSS of the 1S$_{\text{e}}$ states is then
\begin{align}\label{eq:OSS_e}
\delta E^c_{00\beta}=&
\frac{1}{2}\frac{e^2 |F|^2}{\epsilon_0 n_{\text{s}} c} I_0 \tilde{\Delta}_{1\text{S}_{\text{e}}, 1\text{S}_{3/2}}^{-1} \nonumber\\
&\times 
\displaystyle\sum_{M=-3/2}^{3/2}|\bra{\Psi^c_{00\beta}}\mathbf{e}\cdot\hat{\bold{r}}\ket{\Psi^{v\,+}_{\frac{3}{2}\,M}}|^2.
\end{align}
Similarly, the shift of the hole state $\Psi^{v\,+}_{\frac{3}{2}\,M}$ is given by
\begin{align*}\label{eq:OSS_h}
\delta E^v_{\frac{3}{2}\,M}=&
-\frac{1}{2}\frac{e^2 |F|^2}{\epsilon_0 n_{\text{s}} c} I_0 \tilde{\Delta}_{1\text{S}_{\text{e}}, 1\text{S}_{3/2}}^{-1} \nonumber\\
&\times
\displaystyle\sum_{\beta} |\bra{\Psi^c_{00\beta}}\mathbf{e} \cdot\hat{\bold{r}}\ket{\Psi^{v\,+}_{\frac{3}{2}\,M}}|^2. 
\end{align*}
For light polarized linearly at an angle $\theta$ relative to the crystalline $c$ axis, the wave functions of Eqs.~\ref{eq:electrons} and \ref{eq:holes} yield
\begin{equation}\label{eq:electron_OSS}
\displaystyle\sum_{M=-3/2}^{3/2}|\bra{\Psi^c_{00\beta}}\mathbf{e}\cdot \hat{\bold{r}}\ket{\Psi^{v\,+}_{\frac{3}{2}\,M}}|^2
=\frac{2}{3}\frac{1}{m_0^2\omega_{1\text{S}_{\text{e}}, 1\text{S}_{3/2}}^2}K_0P^2,
\end{equation}
where 
\begin{equation*}
P\equiv P_x=-i\bra{s}\hat{p}_x\ket{x}=-i\bra{s}\hat{p}_y\ket{y}=-i\bra{s}\hat{p}_z\ket{z}
\end{equation*}
is the Kane interband matrix element and $K_0$ is the squared magnitude of the radial overlap integral:
\begin{equation*}
K_0=\left|A_{00}\int_{r=0}^a dr\,r^2\,R_0^+(r) j_0(k_{00}r)\right|^2.
\end{equation*}
Although the shift of the electron states is orientation-independent, the valence band states undergo orientation-dependent Stark shifts, since
\begin{widetext}
\begin{align}\label{eq:hole_OSS}
\displaystyle\sum_{\beta} & |\bra{\Psi^c_{00\beta}}\mathbf{e} \cdot\hat{\bold{r}}\ket{\Psi^{v\,+}_{\frac{3}{2}\,M}}|^2 =\frac{1}{m_0^2\omega_{1\text{S}_{\text{e}}, 1\text{S}_{3/2}}^2}K_0P^2
\times  \begin{cases}
    \frac{1}{2}\sin^2\theta,       & \quad \text{if } M=\pm\frac{3}{2} \\
    \frac{2}{3}\cos^2\theta + \frac{1}{6}\sin^2\theta,  & \quad \text{if } M=\pm\frac{1}{2}
  \end{cases}.
\end{align}
\end{widetext}
Using Eqs.~\ref{eq:OSS_e}--\ref{eq:hole_OSS} in Eq.~\ref{eq:fullOSS}, we find
\begin{align}\label{eq:OSSobs1S}
\delta E^{\text{obs}}_{1\text{S}_{\text{e}},1\text{S}_{3/2}} &=\frac{8}{15} \frac{e^2|F|^2}{\epsilon_0 n_{\text{s}} c} \frac{1}{m_0^2\omega_{1\text{S}_{\text{e}}, 1\text{S}_{3/2}}^2} I_0 \tilde{\Delta}_{1\text{S}_{\text{e}}, 1\text{S}_{3/2}}^{-1} K_0P^2\nonumber\\
&=\gamma_0 \frac{|F|^2}{\epsilon_0 n_{\text{s}} c} I_0 
\tilde{\Delta}_{1\text{S}_{\text{e}}, 1\text{S}_{3/2}}^{-1}
\nonumber\\
&\qquad\times
\sum_{\beta,M}  |\bra{\Psi^c_{00\beta}}\mathbf{e} \cdot\hat{\boldsymbol{\mu}}\ket{\Psi^{v\,+}_{\frac{3}{2}\,M}}|^2,
\end{align}
which is Eq.~\ref{eq:OSS1S} with $\gamma(d,E_p)=\gamma_0\equiv2/5$.

\begin{table*}[t!]
\caption{\label{tab:OSS}Calculated contributions to the 1S$_{3/2}$1S$_e$ oscillator strength. The primary contributions to $\gamma$ from transitions involving the 1S$_{\text{e}}$ state are shown for CdSe NQDs of different diameter under varying Stark pump detuning. The total value of $\gamma(d,E_p)$ is the sum of the $\gamma_{i,j}$ shown as well as the calculated contributions from intraband hole transitions and minor contributions from other electron transitions. The measured values of $S$, the observed OSS versus $I_0\tilde{\Delta}_{1\text{S}_{\text{e}}, 1\text{S}_{3/2}}^{-1}$, and the square of the calculated local field factor, $|F|^2$, are used in calculating $f_{1\text{S}_{3/2}1\text{S}_{\text{e}}}$ from $\gamma$. Within a given row, the difference between $\gamma$ and the sum of the $\gamma_{i,j}$ is primarily due to the sum of $\gamma_{1\text{S}_{3/2},j}$ over the various hole states, $j$.} 
\resizebox{\textwidth}{!}{%
\begin{tabular}{| c | c | c | c | c | c | c | c | c | c | c | c | c | c | c |}
\hline
\parbox{1.3cm}{\centering$E_{1\text{S}_{3/2}1\text{S}_{\text{e}}}$ \bigstrut \\  (eV)} 
& \parbox{1.3cm}{\centering diameter \\ (nm)\cite{Jasieniak2009} }
& \parbox{1.0cm}{\centering$E_p$ \\ (eV)} 
& $\gamma_{1\text{S}_{3/2},1\text{S}_{\text{e}}}$ 
 & $\gamma_{1\text{S}_{\text{e}},2\text{S}_{3/2}}$ 
 & $\gamma_{1\text{S}_{\text{e}},1\text{S}_{1/2}}$ 
 & $\gamma_{1\text{S}_{\text{e}},2\text{S}_{1/2}}$ 
 & $\gamma_{1\text{S}_{\text{e}},3\text{S}_{3/2}}$ 
 & $\gamma_{1\text{S}_{\text{e}},1\text{P}_{\text{e}}}$
& $\gamma$ 
& $|F|^2$ 
& \parbox{1.8cm}{\centering S \\ (meV GW cm$^{-2}$ eV$^{-1}$) } 
& $f_{1\text{S}_{3/2}1\text{S}_{\text{e}}}$ \\ \hline
2.455 & 2.493 & 2.156  & 0.40 & 0.023 & 0.041 & 0.015 & 0.00 & 0.087 & 0.62 & 0.412 & 0.264 & 10.7\\ \hline
2.332 & 2.865 & 2.073  & 0.40 & 0.027 & 0.035 & 0.018 & 0.001 & 0.083 & 0.62 & 0.402 & 0.247 & 10.3\\ \hline
2.181 & 3.616 & 1.904  & 0.40 & 0.032 & 0.015 & 0.036 & 0.000 & 0.011 & 0.66 & 0.389 & 0.310 & 12.5\\ \hline
2.049 & 4.838 & 1.784  & 0.40 & 0.031 & 0.001 & 0.039 & 0.000 & 0.13 & 0.65 & 0.376 & 0.297 & 12.6\\ \hline
1.997 & 5.577 & 1.784  & 0.40 & 0.032 & 0.000 & 0.036 & 0.002 & 0.11 & 0.61 & 0.371 & 0.330 & 15.0\\ \hline
1.937 & 6.708 & 1.741  & 0.40 & 0.033 & 0.000 & 0.012 & 0.022 & 0.11 & 0.603 & 0.366 & 0.295 & 13.8\\ \hline
2.455 & 2.493 & 1.55  & 0.40 & 0.032 & 0.061 & 0.027 & 0.002 & 0.27 & 1.27 & 0.446 & 0.239 & 10.7\\ \hline
2.332 & 2.865 & 1.55  & 0.40 & 0.036 & 0.053 & 0.032 & 0.001 & 0.41 & 1.18 & 0.435 & 0.232 & 10.9\\ \hline
2.181 & 3.616 & 1.55  & 0.40 & 0.039 & 0.020 & 0.054 & 0.001 & 0.36 & 1.08 & 0.420 & 0.211 & 10.5\\ \hline
2.049 & 4.838 & 1.55  & 0.40 & 0.036 & 0.001 & 0.055 & 0.000 & 0.31 & 0.92 & 0.405 & 0.228 & 13.0\\ \hline
1.997 & 5.577 & 1.55  & 0.40 & 0.038 & 0.000 & 0.055 & 0.003 & 0.28 & 0.85 & 0.399 & 0.225 & 13.7\\ \hline
1.937 & 6.708 & 1.55  & 0.40 & 0.038 & 0.000 & 0.018 & 0.033 & 0.26 & 0.81 & 0.393 & 0.387 & 12.3\\ \hline
\end{tabular}}
\end{table*}

In considering all contributions to $\delta E^{\text{obs}}_{1\text{S}_{\text{e}},1\text{S}_{3/2}}$, we distinguish between interband and intraband transitions. Qualitatively, the primary difference between these terms in calculations of $\delta E^{\text{obs}}_{1\text{S}_{\text{e}},1\text{S}_{3/2}}$ is the transition matrix element in Eq.~\ref{eq:OSS}. In calculating the OSS of an electronic state due to an interband transition $i\rightarrow j$, it is easiest to calculate the matrix element $\vec{\mu}_{ji}$ in Eq.~\ref{eq:OSS} by replacing $\vec{r}_{ji}$ by $-i\vec{p}_{ji}/m_0\omega_{ji}$. In interband transitions, $\vec{p}_{ji}$ yields a matrix element between Bloch wave functions, and the envelope wave functions simply yield the squared magnitude of a radial overlap integral, $K_{i,j}$, in the same way that $K_0$ appears in the first line of Eq.~\ref{eq:OSSobs1S}. Conversely, when calculating the OSS of an electronic state due to an intraband transition, it is easiest to calculate $\vec{r}_{ji}$, which yields a radial matrix element, rather than a radial overlap integral, between the envelope wave functions. For example, when calculating the OSS of the 1S$_{\text{e}}$ state due to an intraband transition, there appears instead of $K_{i,j}$ a term 
\begin{equation*}
\tilde{K}_{1\text{S}_{\text{e}},i}=\left|A^*_i A_{00}\int_{r=0}^a dr\,r^3 j_{l_i}(k_{c,l_i}r)j_0(k_{c,00}r)\right|^2.
\end{equation*}
In $\tilde{K}$ there appears in the integrand a factor of $r^3$, rather than the factor of $r^2$ that appears in the integrand of $K_{i,j}$.

Using the single-particle, effective-mass approach outlined above, we have calculated the OSS due to all transitions making significant contributions to $\delta E^{\text{obs}}_{1\text{S}_{3/2}1\text{S}_{\text{e}}}$. CdSe NQD size was assigned according to the sizing curve of Jasieniak et al.\cite{Jasieniak2009}. Past calculations of the electronic structure of CdSe NQDs using the effective mass approximation have typically assumed an infinite confinement potential\cite{Ekimov1993} or a finite but unrealistically large confinement\cite{NorrisBawendi1996}. We have chosen an electron confinement potential of 4~eV based on calculations of valence band offsets\cite{WeiZunger1998} and experimental cyclic voltammetry and one- and two-photon photoemission\cite{Markus2009}. In general, high-energy states are expected to have highly oscillatory wave functions and so will have small radial integrals with the 1S$_{\text{e}}$ and 1S$_{3/2}$ states. Therefore, for all types of transitions, we set cutoffs for the energies of the transitions that we consider. These cutoffs correspond to electron confinement energies greater than about 1.6~eV, hole confinement energies greater than about 0.8~eV, and interband transition energies greater than 3.5 eV.

$\gamma(d,E_p)$ can be expressed as a sum of individual terms $\gamma_{i,j}(d,E_p)$ corresponding to the contribution of transition $i\rightarrow j$ to $\delta E^{\text{obs}}_{1\text{S}_{3/2}1\text{S}_{\text{e}}}$. The interband contributions depend on the electron and hole radial functions ($j_i(r)$ and $R^{\pm}_l(r)$, respectively) through the radial overlap integral $K^{eh}_{i,k}$. Since the radial functions do not undergo large changes in shape with respect to $r/a$, these contributions are not expected to depend very sensitively on details such as the magnitudes of the confinement potentials; for a 4~eV electron confinement potential, the radial integral for the 1S$_{3/2}$1S$_{\text{e}}$ transition ranges from 0.61 in our smallest NQDs to 0.76 in our largest. This can be compared to a nearly size-independent value of 0.93 for an infinite electron confinement potential.

We consider the electron and hole intraband transitions separately. The electron intraband contributions to the OSS depend on the electron radial functions via $\tilde{K}^{ee}_{1\text{S}_{\text{e}},j}$ above, which leads to a $\sim a^2$ dependence of $K^{ee}_{1\text{S}_{\text{e}},j}$ on NQD radius, $a$. The far largest such integral is for the 1S$_{\text{e}}\rightarrow$1P$_{\text{e}}$ transition. In the case of an infinite confinement potential, $|\int_0^a dr\,r^3 f_{1\text{S}_{\text{e}}}(r)f_{1\text{P}_{\text{e}}}(r)|^2=0.28a^2$\cite{GuyotSionnest1998}. For a 4~eV confinement potential, the electron wave function has a greater amplitude at larger $r$ compared to the case of infinite confinement, so that the integral for a 4~eV confinement potential is $\sim$50\% larger than for an infinite confinement potential in the largest dots and $\sim$85\% larger in the smallest dots. However, in the OSS, the growth in the radial integral for intraband transitions is substantially offset by the diameter-dependence of the intraband transition energies. The intraband transition energies scale as $\sim a^{-2}$, so that for large, blue detunings of the Stark pump from the intraband transitions, $\tilde{\Delta}_{j,1\text{S}_{\text{e}}}^{-1}$ scales approximately as $a^{-2}$ and the product $\tilde{K}^{ee}_{1\text{S}_{\text{e}},j} \tilde{\Delta}_{j,1\text{S}_{\text{e}}}^{-1}$ is only weakly size-dependent. 

We treat the hole intraband transitions somewhat differently than the electron intraband transitions. The holes are markedly heavier than the electrons and so are more sensitive to variations in the potential energy across the dot, including surface variations, core defects, and room-temperature phonon populations. Notably, the Coulomb energy is expected to exceed the hole confinement energy in all but the two smallest NQDs we have studied. This is in contrast to the electrons for which, despite the use of a finite confinement potential in our calculations, the electron confinement energy dominates the Coulomb energy for all dot sizes studied here. Our approach is then to use the values of the hole-hole radial integrals, $\tilde{K}$, for the NQDs with smallest diameter in calculating the contributions in all dots of the hole intraband transitions to $\delta E^{\text{obs}}_{1\text{S}_{3/2}1\text{S}_{\text{e}}}$. In this case, changes of the intraband contributions with dot size are due to the change in energy spacings, which are estimated experimentally, where possible, or calculated. 

\begin{figure}[t!]
\centering
	\includegraphics[width=\linewidth]{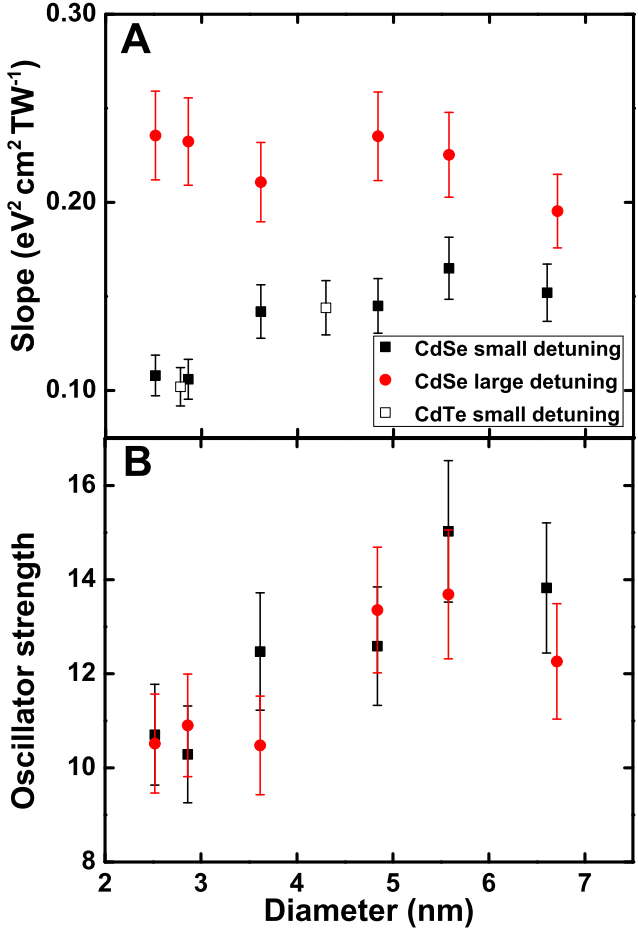}
	\caption{\label{fig:slopes}\textbf{Optical Stark slopes and oscillator strengths.} 
Panel (\textbf{A}) shows the slope of the OSS, i.e., $\delta E^{\text{obs}}_{1\text{S}_{\text{e}}1\text{S}_{3/2}}$ versus $I_0 \tilde{\Delta}_{1\text{S}_{\text{e}}, 1\text{S}_{3/2}}^{-1}$, for CdSe and CdTe NQDs of different size. ``Small detuning\rq\rq{} refers to detunings for which $E_{1\text{S}_{\text{e}}, 1\text{S}_{3/2}}-E_p<0.3$~eV in Table~\ref{tab:OSS}. The slopes and values of $\gamma$ from Table~\ref{tab:OSS} yield the oscillator strengths for CdSe NQDs shown in panel (\textbf{B}).}
\end{figure}	

For calculations of the local field factor, we assume spherical NQDs, for which 
\begin{equation*}
F=\frac{3\epsilon_s}{\epsilon_{\text{NQD}}+2\epsilon_s},
\end{equation*}
where $\epsilon_s$ and $\epsilon_{\text{NQD}}$ are the relative permittivities of the solvent and NQD, respectively. For energies below the 1S$_{3/2}\rightarrow$1S$_\text{e}$ transition, we use $\epsilon_{\text{NQD}}=\epsilon_{\infty,\text{pp}}+\delta\epsilon_{\text{res}},$ where $\epsilon_{\infty,\text{pp}}$ is the size-dependent high-frequency, non-resonant relative permittivity obtained from pseudopotential calculations\cite{WangZunger1996} and where $\delta\epsilon_{\text{res}}$ is a size-independent resonant electronic contribution to the permittivity equal to $\epsilon_{\text{bulk}}\left(\hbar\omega\right) -\epsilon_{\infty,\text{pp,bulk}}$, and  $\epsilon_{\text{bulk}}\left(\hbar\omega\right)$ is the measured relative permittivity for bulk CdSe\cite{Logothetidis1986}. 

The calculated values of $\gamma$, the dominant contributions $\gamma_{i,j}$, and $|F|^2$ are presented in Table~\ref{tab:OSS}. Also shown in Table~\ref{tab:OSS} and Fig.~\ref{fig:slopes}A is the slope, $S$, obtained from linear fits to the experimentally measured values of $\delta E^{\text{obs}}_{1\text{S}_{3/2}1\text{S}_{\text{e}}}$  versus $I_0\tilde{\Delta}_{1\text{S}_{\text{e}}, 1\text{S}_{3/2}}^{-1}$ shown in Fig.~\ref{fig:size_exc_dep}A, as well as the corresponding data for a Stark pump of 1.55~eV (not shown). Our values of $\gamma$ and $S$ yield the oscillator strengths of the 1S$_{3/2}\rightarrow$1S$_{\text{e}}$ transition shown in Table~\ref{tab:OSS} and Figs.~\ref{fig:size_exc_dep}B and \ref{fig:slopes}B. In particular, we note that despite the different OSS measured for moderate ($-0.2 - -0.3$~eV) and large ($-0.4 - -0.9$~eV) detunings shown in Fig.~\ref{fig:slopes}A, the values of the oscillator strength $f_{1\text{S}_{3/2} 1\text{S}_{\text{e}}}$ determined from the two sets of measurements are in agreement.

Also shown in Fig.~\ref{fig:slopes}A are the OSS slopes for a pair of CdTe samples comparable in size (we use the CdTe sizing curves given by Yu et al.\cite{Yu2003}) to CdSe NQDs in the smallest and intermediate size ranges. The CdTe OSS data are very similar to the CdSe data, as expected given the similar bulk parameters (bandgap, Kane parameter, and electron and hole effective masses). Given that for CdTe NQDs of 3--7~nm diameter $f^{\text{QD}}_{1\text{S}_{3/2} 1\text{S}_{\text{e}}}\approx11$\cite{KamalHens2012}, one would again expect an oscillator strength in strongly confined CdSe NQDs  
similar to what we have obtained. 

\section{Comparisons to previous reports}
The derivation of the oscillator strength of the 1S$_{3/2}$1S$_{\text{e}}$ excitons, $f_{1\text{S}_{3/2}1\text{S}_{\text{e}}}$, of previous studies (Fig.~\ref{fig:size_exc_dep}B) is based on the reported peak value, $\sigma(\hbar\omega_{1\text{S}_{\text{e}},1\text{S}_{3/2}})$, and the corresponding energy half-width-half-maximum, $\Delta_{\text{HWHM}}$, determined from the spectrum of the absorption cross section per NQD. Combined with Eq.~\ref{eq:sigma1S}, these yield  
\begin{align*}
\sigma(\hbar\omega_{1\text{S}_{\text{e}}, 1\text{S}_{3/2}}) =& \frac{\sqrt{\ln 2}}{\Delta_{\text{HWHM}}\sqrt{\pi}} \\
&\quad\times\frac{\pi e^2\hbar|F(\hbar\omega_{1\text{S}_{\text{e}}, 1\text{S}_{3/2}})|^2}{2\epsilon_0 n_{\text{s}}m_0c}f_{1\text{S}_{3/2}1\text{S}_{\text{e}}}.
\end{align*}
Likewise, this equation yields  $\sigma(\hbar\omega_{1\text{S}_{\text{e}}, 1\text{S}_{3/2}})$ from our values of $f_{1\text{S}_{3/2}1\text{S}_{\text{e}}}$ and our measured 1S$_{3/2}$1S$_{\text{e}}$ linewidths. In previous studies \cite{Jasieniak2009,Leatherdale2002,Klimov2000,Yu2003}, the sizing curve affects the value of the oscillator strength. Therefore, for analyzing all studies we corrected the oscillator strength by multiplying by a factor of $(d^\ast/d)^3$ where $d^\ast$ and $d$ are the corrected (from Ref.~\cite{Jasieniak2009}) and originally reported sizes, respectively. In Fig.~\ref{fig:abs_spec}B, we convert $f_{1\text{S}_{3/2}1\text{S}_{\text{e}}}$ to $\sigma(\hbar\omega_{1\text{S}_{\text{e}}, 1\text{S}_{3/2}})$ in a similar way and scale the absorption spectrum to match $\sigma(\hbar\omega_{1\text{S}_{\text{e}}, 1\text{S}_{3/2}})$.

\section{\label{app:sat}DA saturation measurements of the absorption cross section}

\begin{figure}[t!]
\centering
	\includegraphics[width= \linewidth]{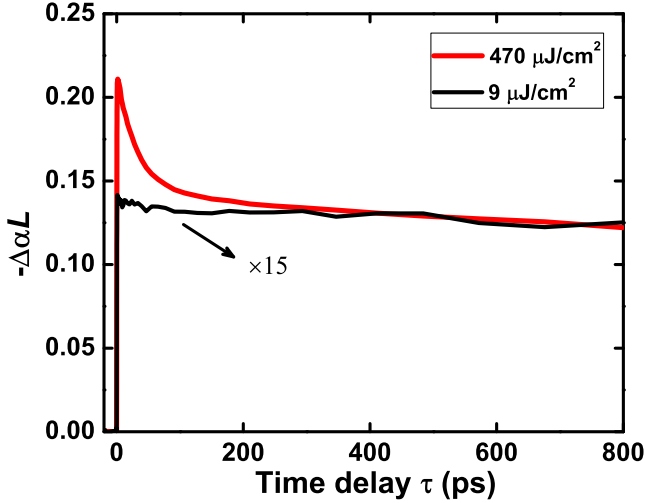}
	\caption{\label{fig:TA_dynamics}\textbf{DA dynamics of the 1S$_{
\bold{3/2}}$1S$_\mathbf{e}$ peak of 3.6~nm CdSe NQDs.} The samples are pumped at 3.1~eV. The red and black solid curves are the DA dynamics under fluences of 470 and 9 $\mu$J/cm$^2$ per pulse, respectively.	The low-fluence data are scaled to highlight the absence of multi-exciton Auger recombination and to show the common dynamics at long delays for both low- and high-fluence excitation.}
\end{figure}

In Fig.~\ref{fig:TA_dynamics}, we show a pair of representative DA dynamics traces for 3.6~nm CdSe NQDs pumped at 3.1~eV (i.e., $>0.6$~eV above the 1S$_{3/2}$1S$_{\text{e}}$ absorption peak) and probed at the 1S$_{3/2}$1S$_{\text{e}}$ transition. At low intensity, the samples are in the single-exciton regime for the entire delay range. At high intensity, the samples experience Auger recombination of multi-exciton states before relaxing into the single-exciton regime at $t>200$~ps. When normalized at long delays to account for the different signal levels, we see that the dynamics of strongly excited samples after Auger recombination are the same as the dynamics of single-excitons as reflected under low-fluence excitation. In particular, the amplitude of the signal at long delays should depend only on the probability that at least one pump photon was absorbed and the bleach induced by the single electron or exciton that remains after Auger recombination. 

\begin{figure}[t!]
\centering
	\includegraphics[width=\linewidth]{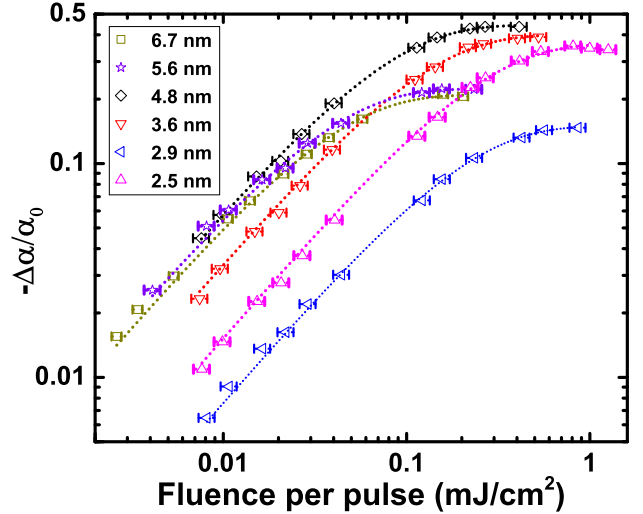}
	\caption{\label{fig:sat_abs}\textbf{Saturation measurements of the absorption cross section at 3.1~eV.} The fluence-dependence of $-\Delta\alpha/\alpha_0$ at the 1S$_{3/2}$1S$_{\text{e}}$ peak measured at $\tau=800$~ps for samples with diameter from 2.5~nm to 6.7~nm pumped at 3.1~eV. The dotted curves are fits with the function $-\Delta\alpha\left(\Phi\right)/\alpha_0=A\left\{1-\exp\left[ -\sigma\left(3.1\,\text{eV}\right) \Phi\right]\right\}$, where the fit parameters $A$ and $\sigma\left(3.1\,\text{eV}\right)$ are respectively the saturated value of $-\Delta\alpha/\alpha_0$ and the absorption cross section per NQD at 3.1~eV.}
\end{figure}

DA saturation curves of the CdSe 1S$_{3/2}$1S$_\text{e}$ peak at $t=0.8$~ns are shown in Fig.~\ref{fig:sat_abs}. No matter how many excitons are initially excited, no more than one exciton remains in each NQD by $\tau=800$~ps. Therefore, at long delays $\sigma\left(3.1\text{ eV}\right)$  can be determined by fitting to the Poisson probability of absorption of at least one pump photon per dot: 
\begin{equation*}
\frac{-\Delta \alpha}{\alpha_0}=A\left[1-P_0\left(\sigma\Phi\right)\right]=A\left[1-\exp(-\sigma \Phi)\right]),
\end{equation*}
where $P_0\left(\sigma\Phi\right)$ is the probability of an NQD absorbing zero photons, $\Phi$ is the pump fluence, and the fitting parameters $\sigma$ and $A$ are the absorption cross section per dot at 3.1~eV and the saturated DA signal, respectively. 

\section{\label{app:excitonOSS}1S$_{3/2}$1S$_{\text{e}}$ Stark shift: exciton picture}
Here we show how the energy-integrated oscillator strength of the 1S$_{3/2}$1S$_{\text{e}}$ peak is affected by the size-dependent 1S$_{3/2}$1S$_{\text{e}}$ exciton fine structure. For simplicity, we consider only the contribution of the 1S$_{3/2}\rightarrow$1S$_{\text{e}}$ transition to the OSS of the 1S$_{3/2}$1S$_{\text{e}}$ peak.  

The OSS of a transition from the ground-state, $G$, to a given exciton, $X_{n}$, is determined by the shift of $X_n$ due to the OSE through the $G\rightarrow X_n$ transition, the shift of $G$ due to the OSE through \textit{all} transitions out of the ground state,  and the shift of $X_n$ due to transitions from $X_n$ to all optically allowed biexcitons, $XX_k$ \cite{Combescot1989}:
\begin{align*}
	\delta E_{X_n,G}=& \frac{1}{2}\frac{e^2|F|^2}{\epsilon_0 n_{\text{s}} c} I_0 
\left\{\vphantom{\displaystyle\sum_{k}}|\bra{X_n}\mathbf{e}\cdot\hat{\bold{r}}\ket{G}|^2
\tilde{\Delta}_{X_n,G}^{-1} \right. \\
& \quad +\displaystyle\sum_{j}|\bra{X_j}\mathbf{e}\cdot\hat{\bold{r}}\ket{G}|^2
\tilde{\Delta}_{X_j,G}^{-1} \\
&\quad \left.+ \displaystyle\sum_{k}
|\bra{XX_k}\mathbf{e}\cdot\hat{\bold{r}}\ket{X_n}|^2 \tilde{\Delta}_{XX_k,X_n}^{-1} \right\} \nonumber
\end{align*}
In the case that the fine structure splittings and biexciton binding are small compared to the Stark field detuning, we can approximate $\tilde{\Delta}_{XX_k,X_n}^{-1}\approx\tilde{\Delta}_{X_j,G}^{-1} \approx \tilde{\Delta}_{X_n,G}^{-1},$ so that
\begin{align}
\label{eqn:biex_eff}
	\delta E_{X_n,G}=& \frac{1}{2}\frac{e^2|F|^2}{\epsilon_0 n_{\text{s}} c} I_0
\tilde{\Delta}_{X_n,G}^{-1}
\left( \vphantom{\displaystyle\sum_{k}}
|\bra{X_n}\mathbf{e}\cdot\hat{\bold{r}}\ket{G}|^2 \right.\nonumber\\
&
+\displaystyle\sum_{j}|\bra{X_j}\mathbf{e}\cdot\hat{\bold{r}}\ket{G}|^2  \nonumber\\
&\left.+ \displaystyle\sum_{k}
|\bra{XX_k}\mathbf{e}\cdot\hat{\bold{r}}\ket{X_n}|^2  \right).
\end{align}

In addressing the exciton fine structure, we follow the treatment of Efros et al.\cite{Efros1996} The four 1S$_{3/2}$ and two 1S$_{\text{e}}$ states give rise to eight exciton states (the exciton fine structure). These are labeled by the projection of their total angular momentum along the $c$ axis: the bright (dipole-allowed) states $X=0^{U}$, $1^{\pm U}$, and $1^{\pm L}$ and the dark states $X=0^L$ and $X=\pm2$. The exciton wave functions can be expressed as products of radial envelope functions and Bloch functions of the conduction and valence bands. Using single-particle, electron($e$)-hole($h$) basis functions $\ket{e,h}$ and dropping explicit reference to the radial functions and the total angular momentum of the single-particle states, the exciton states can be written
\begin{align*}
	\ket{0^{U}}&=\frac{1}{\sqrt{2}}\Big(-i\ket{\uparrow;-1/2}+\ket{\downarrow;+1/2}\Big) \\
	\ket{+1^{U,L}}&=\mp iC_{\pm}\ket{\uparrow;+1/2}+C_{\mp}\ket{\downarrow;+3/2} \\
	\ket{-1^{U,L}}&= \mp iC_{\mp}\ket{\uparrow;-3/2} + C_{\pm}\ket{\downarrow;-1/2}
\end{align*}
where $\ket{\beta;m}$ refers to the electron-hole pair state consisting of a spin $\beta$ electron in the conduction band and a $\ket{J=3/2,m}$ hole, the upper(lower) sign of the $\pm$ and $\mp$ pairs are associated with the $U$($L$) states, \begin{equation*}
C_{\pm}=\sqrt{\frac{\sqrt{\psi^2+3\eta^2}\pm \psi}{2\sqrt{\psi^2+3\eta^2}}},
\end{equation*}
$\psi=(\Delta-2\eta)/2$, $\Delta$ is the total splitting of the hole state and is the sum of crystal field and shape splitting, and $\eta$ is given by
\begin{equation*}
\eta=\left(\frac{a_{\text{B}}}{a}\right)^3\hbar\omega_{\text{ST}}\chi\left(\beta\right),
\end{equation*}
where $a_{\text{B}}=5.6$~nm is the bulk exciton Bohr radius, $\hbar\omega_{\text{ST}}=0.13$~meV is the singlet-triplet splitting of the lowest-energy exciton in bulk, $\beta=0.28$ is the ratio of heavy-hole to light-hole masses, and $\chi\left(\beta\right)$ describes the radial overlap of the electron and hole envelope functions. 

For light with linear polarization at an angle $\theta$ to the NQD $c$ axis, the squared magnitude of the momentum matrix elements, $p_{X_n,G}$, between the ground state ($G$) and the bright 1S$_{3/2}$1S$_{\text{e}}$ excitons ($X_n$) are then given by
\begin{align*}
|p_{0^U,G}|^2&=\frac{4}{3} K_0P^2 \cos^2\theta \\
|p_{-1^{U,L},G}|^2&=|p_{+1^{U,L},G}|^2 \\
	&=\frac{1}{6}\left(1+2C_{\mp}^2 \pm \frac{3\eta}{\sqrt{\psi^2+3\eta^2}}\right) K_0P^2 \sin^2\theta.
\end{align*}

To calculate the impact of biexcitons (the third term in Eq.~\ref{eqn:biex_eff}) in the OSS we follow the treatment by Rodina and Efros of the biexciton fine structure derived from the 1S$_{3/2}$ and 1S$_{\text{e}}$ states \cite{Rodina2010}. There are only six possible biexciton states derived from the 1S$_{3/2}$ and 1S$_{\text{e}}$ states: a four-fold degenerate set of states with total angular momentum $N=2$ and labeled by the projection of angular momentum onto the $c$ axis of $M_N=\pm2$ and $M_N=\pm1$, and two non-degenerate states of $M_N=0$, labeled $0^+$ and $0^-$. The relative probabilities for transitions from each of the 1S$_{3/2}$1S$_{\text{e}}$ excitons to each of the 1S$_{3/2}$1S$_{\text{e}}$ biexcitons is given in Table~1 of Ref.~\onlinecite{Rodina2010}. For example, the sum of squared momentum matrix elements for transitions from $0^U$ to the various biexciton states is
\begin{widetext}
\begin{align*}
	\sum_{i}|p_{XX_i,0^U}|^2&=\left|\bra{XX_{0^{-}}}
\mathbf{e}\cdot\hat{\bold{p}}\ket{0^U}\right|^2 
+\left|\bra{XX_{0^+}}
\mathbf{e}\cdot\hat{\bold{p}}\ket{0^U}\right|^2 
+\left|\bra{XX_{+1}}\mathbf{e}\cdot\hat{\bold{p}}\ket{0^U}\right|^2  
+\left|\bra{XX_{-1}}
\mathbf{e}\cdot\hat{\bold{p}}\ket{0^U}\right|^2\\
&
=\left(\frac{1}{2}\sin^2\theta +\frac{4}{3}\cos^2\theta\right) K_0P^2.
\end{align*}
Similarly, the momentum matrix elements for the biexciton transitions from the other bright single excitons yield
\begin{align*}
\sum_{i}&|p_{XX_i,\pm1^{U}}|^2 
=\frac{1}{3}\left(2-2C_+^2 \cos^2\theta +\sqrt{3}C_+C_-\sin^2\theta \right) K_0P^2 \\
\sum_{i}&|p_{XX_i,\pm1^{L}}|^2 
=\frac{1}{3}\left(2-2C_-^2 \cos^2\theta - \sqrt{3}C_+C_-\sin^2\theta \right) K_0P^2.
\end{align*}
\end{widetext}

With the relative probabilities for all of the ground-to-single-exciton transitions and single-to-biexciton transitions, we can calculate the OSS for each of the transitions from the ground to single-exciton states associated with the 1S$_{3/2}$1S$_{\text{e}}$ peak. The OSS of the ground state is given by
\begin{align*}
\delta E_{G}&=-\frac{1}{2}\xi\left(2|p_{+1^U,G}|^2+2|p_{+1^L,G}|^2+|p_{0^U,G}|^2\right)\\
&
=-\frac{2}{3}\xi K_0P^2,
\end{align*}
where, from Eq.~\ref{eqn:biex_eff} and the relationship $\vec{p}_{kj}=im_0\omega_{kj}\vec{r}_{kj}$,
\begin{equation*}
\xi \equiv \frac{1}{m_0^2
\omega_{1\text{S}_{\text{e}},1\text{S}_{3/2}}^2}\frac{e^2|F|^2}{\epsilon_0 n_{\text{s}} c}I_0\tilde{\Delta}_{1\text{S}_{\text{e}}, 1\text{S}_{3/2}}^{-1}.
\end{equation*} 
The shift of the ground state is independent of orientation. The OSSs of the transitions from $G$ to the bright 1S$_{3/2}$1S$_{\text{e}}$ excitons are given by
\begin{align*}
\delta E_{\pm 1^U,G}&=\frac{1}{2}\xi \left(|p_{\pm 1^U}|^2-|p_{\pm XX,1^U}|^2\right)-\delta E_{G} \\
\delta E_{\pm 1^L,G}&=\frac{1}{2}\xi \left(|p_{\pm 1^L}|^2-|p_{XX,\pm 1^L}|^2\right)-\delta E_{G} \\
\delta E_{0^U,G}&=\frac{1}{2}\xi \left(|p_{0^U}|^2-|p_{XX,0^U}|^2\right)-\delta E_{G}
\end{align*}

Since the 1S$_{3/2}$1S$_{\text{e}}$ absorption peak is broad compared to the splittings between the excitons, then as shown in the Appendix~\ref{app:obsOSS}, the experimentally observed OSS is an average of the OSS of all the transitions that comprise the 1S$_{3/2}$1S$_{\text{e}}$ peak weighted by the relative strength of each transition: 
\begin{widetext}
\begin{align}\label{eq:OSSobs2}
\delta E^{\text{obs}}_{1\text{S}_{\text{e}},1\text{S}_{3/2}}&= \frac{\left\{\displaystyle\int_{0}^{\pi}\left(2\delta E_{+1^U,G} |p_{+1^U,G}|^2 +2\delta E_{+1^L,G} |p_{+1^L,G}|^2 + \delta E_{0^U,G} |p_{0^U,G}|^2\right) \sin\theta d\theta \right\}}{\left\{\displaystyle\int_{0}^{\pi}\left(2|p_{+1^U,G}|^2+2|p_{+1^L,G}|^2 +|p_{0^U,G}|^2\right)\sin\theta d\theta\right\}} \nonumber \\
&=\frac{8}{15}\xi K_0P^2 \left\{1-\frac{1}{12}\left[C_{+}^2C_{-}^2 - \frac{\sqrt{3}}{2}\left(C_-^2-C_+^2\right)C_+C_-\right] \right\}. 
\end{align}
\end{widetext}

\begin{figure}[t!]
\centering
	\includegraphics[width=\linewidth]{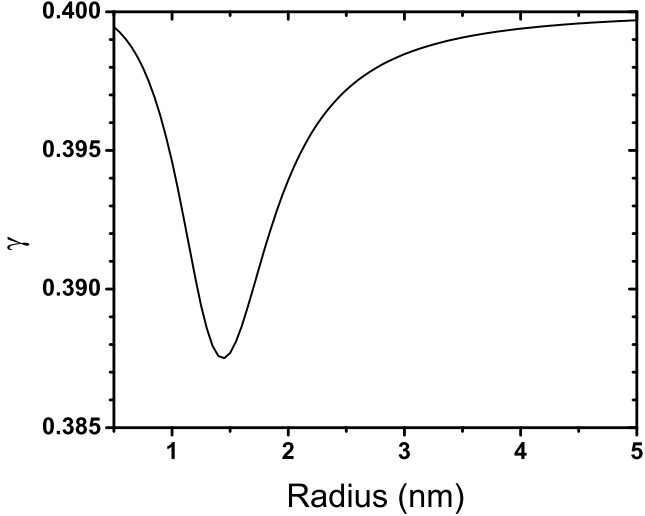}
	\caption{\label{fig:gamma_vs_R}\textbf{Size-dependent scaling factor $\gamma$} relating the magnitude of the observed optical Stark shift of the 1S$_{3/2}$1S$_{\text{e}}$ peak to the total oscillator strength of the 1S$_{3/2}$1S$_{\text{e}}$ peak.}
\end{figure}		

Noting that angular averaging of $|\mathbf{e}\cdot\vec{\mu}_{n,G}|^2$ over all orientations yields
\begin{align}
\label{eq:dipole_squared}
\sum_{n}\left<|\mathbf{e}\cdot\vec{\mu}_{n,G}|^2\right>
&=e^2\displaystyle\sum_{\beta,M}|\bra{\Psi^c_{00\beta}}\mathbf{e}\cdot \hat{\bold{r}}\ket{\Psi^{v\,+}_{\frac{3}{2}\,M}}|^2 \nonumber\\
&=\frac{e^2}{m^2 \omega_{1\text{S}_{\text{e}},1\text{S}_{3/2}}^2} \frac{4}{3}K_0P^2,
\end{align} 
we can rewrite Eq.~\ref{eq:OSSobs2} as
\begin{align}\label{eq:OSSobs}
\delta E^{\text{obs}}
_{1\text{S}_{\text{e}},1\text{S}_{3/2}}=&\gamma\frac{|F|^2}{\epsilon_0 n_{\text{s}} c}\frac{I_0} 
{\tilde{\Delta}_{1\text{S}_{\text{e}}, 1\text{S}_{3/2}}}\sum_n \left<|\bra{X_n}\mathbf{e}\cdot\hat{\boldsymbol{\mu}}\ket{G}|^2\right>,
\end{align}
where
\begin{align*}
\gamma&\equiv \frac{2}{5}\left\{1-\frac{1}{12}\left[C_{+}^2C_{-}^2 - \frac{\sqrt{3}}{2}\left(C_-^2-C_+^2\right)C_+C_-\right] \right\}. 
\end{align*}
Using the fact that $|C_+C_-|^2<1/4$ and $|\left(C_-^2-C_+^2\right)C_+C_-|\leq1/4$, we find that $\gamma\approx 2/5$, which is the value of $\gamma_0$ found in the single-particle picture. In Figure \ref{fig:gamma_vs_R}, we plot $\gamma$, which shows deviations of $<4$\% from the value of $\gamma_0=2/5$. In other words, for large detunings, $\gamma$ is nearly equal to the size-independent single-particle value.

\bibliography{C:/Users/JAM/pubs/CdSe/OSE_lib_JAM}

\end{document}